\documentclass[12pt]{article}
\usepackage[dvips]{graphicx}
\def\be{\begin{equation}}					 
\def\ee{\end{equation}}
\def\ber{\begin{eqnarray}}
\def\eer{\end{eqnarray}}	
\def\dint{\mathop{\intop\kern-0.5em\intop}}
\def\ovc#1{\displaystyle\mathop{#1}^{\kern0.2em\circ}}

\begin{document}
\vspace*{1cm}
\begin{center}
{\Large \bf ${\cal L}$-Particle and Kaluza-Klein World}\\

\vspace{4mm}

{\large A.A. Arkhipov\\
{\it State Research Center ``Institute for High Energy Physics" \\
 142280 Protvino, Moscow Region, Russia}}\\
\end{center}

\vspace{4mm}
\begin{abstract}
{We present some arguments in favour of that the
Kaluza-Klein picture of the world has been confirmed in the
experiments  at very low energies where the nucleon-nucleon dynamics
has been studied. Early predicted ${\cal L}$-particle is related to
the new scale of strong nucleon-nucleon forces in the spirit of
Kaluza-Klein approach. It is shown that KK excitations remarkably
describe the experimentally observed mass spectrum of
diproton system.} 
\end{abstract}

\section{Introduction: Kaluza-Klein Picture}
The original idea of Kaluza and Klein is based on the hypothesis that
the input space-time is a $(4+d)$-dimensional space ${\cal
M}_{(4+d)}$ which can be represented as a tensor product of the
visible four-dimensional world $M_4$ with a compact internal
$d$-dimensional space $K_d$
\be
{\cal M}_{(4+d)} = M_4 \times K_d.   \label{1}
\ee
The compact internal space $K_d$ is space-like one i.e. it has only
spatial dimensions which may be considered as extra spatial
dimensions of $M_4$. In according with the tensor product structure 
of the space ${\cal M}_{(4+d)}$ the metric may be chosen in a
factorizable form. This means that if $z^M = 
\{ x^{\mu}, y^{m}\}$, ($M=0,1,\ldots,3+d,\, \mu = 0,1,2,3,\, m=1,2,
\ldots, d$), are local coordinates on ${\cal M}_{(4+d)}$ then the
factorizable metric looks like
\[
ds^{2} = {\cal G}_{MN}(z) dz^M dz^N =
g_{\mu \nu}(x) dx^{\mu} dx^{\nu} + \gamma_{mn}(x,y) dy^{m} dy^{n},
\]
where $g_{\mu \nu}(x)$ is the metric on $M_4$.

In the year 1921, Kaluza proposed a unification of the
Einstein gravity and the Maxwell theory of electromagnetism in four
dimensions starting from Einstein gravity in five dimensions. He
assumed that the five-dimensional space ${\cal M}_5$ had to be a
product of a four-dimensional space-time $M_4$ and a circle $S_1$:
${\cal M}_5 = M_4 \times S_1$. After that the metric
${\cal G}_{MN}(z)$ in ${\cal M}_5$ can be decomposed  into 10
components describing Einstein gravity tensor
${\cal G}_{\mu\nu}\rightarrow g_{\mu\nu}$,
four components ${\cal G}_{\mu 5}\rightarrow A_{\mu}$ forming an
electromagnetic gauge
field, and one component ${\cal G}_{55}\rightarrow \phi$ representing
a scalar field.
It was shown that the zero mode sector of the Kaluza model (the model
is actually a  five-dimensional Einstein gravity) is equivalent to
the four-dimensional theory which describes the Einstein gravity with
a four-dimensional general coordinate transformations and the Maxwell
theory of electromagnetism with a gauge transformations. In his
original work, Kaluza assumed the zero mode of scalar field had to be
positive constant in order to insure a positive energy.

Recently some models with extra dimensions have been proposed to
attack the electroweak quantum instability of the Standard Model
known as hierarchy problem between the electroweak and gravity
scales. To illustrate the main idea how the hierarchy problem may be
solved in a theory with extra dimensions let us consider the Einstein
$(4+d)$-dimensional gravity with  the action
\[
S_{(4+d)} =\frac{1}{16 \pi G_{(4+d)}} \int d^{4+d}z
\sqrt{-{\cal G}}\,{\cal R} ({\cal M}_{(4+d)}), 
\]
where ${\cal G}=\det|{\cal G}_{MN}|$, and the Ricci scalar curvature
${\cal R}({\cal M}_{(4+d)})$ is defined by the metric ${\cal
G}_{MN}$.
By the mode expanding and integrating over $K_{d}$ one obtains the
four-dimensional action
\[
S_{4} = \frac{1}{16 \pi G_N}\int d^{4}x \sqrt{-g}\,{\cal
R}(M_4) + \mbox{non-zero modes}, 
\]
where $g=\det|g_{\mu\nu}|$, $g_{\mu\nu}$ is Einstein gravity tensor
on $M_4$, and the Ricci scalar curvature ${\cal R}(M_4)$ is defined
by the metric $g_{\mu\nu}$,  $G_N$ is four-dimensional gravitational
Newton constant related with the fundamental constant $G_{(4+d)}$ by
the following equation
\be
G_N = \frac{1}{V_d} G_{(4+d)}, \label{G}
\ee
where $V_d$ is the volume of the compact internal space $K_d$. We can
rewrite Eq. (\ref{G}) in terms of the four-dimensional Planck mass
$M_{Pl} = G_N^{-1/2}$ and a fundamental mass scale of the
$(4+d)$-dimensional gravity  $M^{d+2} = G^{-1}_{(4+d)}$ 
\be
M_{Pl}^{2} = V_d M^{d+2}. \label{M}
\ee
The latter formula is often cited as the reduction formula. If $R$
is a characteristic size of $K_d$ then $V_d\sim R^d$, i.e. we suppose
$V_d=C_dR^d$, where $C_d$ is some constant. Eq.~(\ref{M}) gives
\be
M_{Pl}=C_d^{1/2}M(MR)^{d/2}.\label{M2}
\ee
Now, it's clear that, if the size $R$ of the compact internal space
$K_d$ is large compared to the fundamental length $M^{-1}$, the
Planck mass is much larger than the fundamental gravity scale. Going
further on, if we suppose that the fundamental gravity scale is of
the same order as the electroweak scale, $M\sim 1\,TeV$, then a huge
gap between $M_{Pl}$ and $M_{EW}$ is resulted from the large size of
the internal extra space $K_d$. So, the hierarchy problem replaces by
the hierarchy $R/M^{-1}\sim(M_{Pl}/M)^{2/d}$ and becomes the
problem to explain why the size $R$ of extra space $K_d$ is large.
From
Eq.~(\ref{M2}), assuming that $M\sim 1\,TeV$, it follows
\be
R\sim M^{-1}\left(\frac{M_{Pl}}{M}\right)^{2/d}\sim 10^{32/d} \cdot
10^{-17}\,cm.\label{R}
\ee
or
\[
R^{-1}\sim M\left(\frac{M}{M_{Pl}}\right)^{2/d}\sim 10^{3-32/d}\,GeV.
\]
An exceptional case $d=2$  gives $R\sim 1\,mm$, ($R^{-1}\sim
10^{-4}\,eV$) and this is certainly a quite interesting observation.
In the case $d=1$ we have $R\sim 10^{15}\,cm$, i.e. this case is
excluded by unacceptable large value of $R$. For $d>2$ one obtains
\[
d=3, \qquad R\sim 4.6\, 10^{-7}\,cm,\qquad R^{-1}\sim 20\,eV\ \ \ \
\]
\[
.....................................................................
..........
\]
\[
d=6, \qquad R\sim 2.2\, 10^{-12}\,cm,\qquad R^{-1}\sim 4.6\,MeV
\]

It is obviously that the basic idea of the Kaluza-Klein scenario may
be applied to any model in Quantum Field Theory. As example, let us
consider the simplest case of (4+d)-dimensional model of scalar field
with the action  
\be
S = \int d^{4+d}z \sqrt{-{\cal G}} \left[
\frac{1}{2} \left( \partial_{M} \Phi \right)^2 - 
\frac{m^{2}}{2} \Phi^2 + \frac{G_{(4+d)}}{4!} \Phi^4
\right], 
\label{S}
\ee
where ${\cal G}=\det|{\cal G}_{MN}|$, ${\cal G}_{MN}$ is the metric
on ${\cal M}_{(4+d)} = M_4 \times K_d$, $M_4$ is pseudo-Euclidean
Minkowski space-time, $K_d$ is a compact internal $d$-dimensional
space with the characteristic size $R$. Let $\Delta_{K_{d}}$ be the
Laplace operator on the internal space $K_{d}$, and $Y_{n}(y)$ are
ortho-normalized eigenfunctions of the Laplace operator 
\be
\Delta_{K_{d}} Y_{n}(y) = -\frac{\lambda_{n}}{R^{2}} Y_{n}(y),  
\label{Yn}
\ee
and $n$ is a (multi)index labeling the eigenvalue
$\lambda_{n}$ of the eigenfunction $Y_{n}(y)$. $d$-dimensional torus
$T^{d}$ with equal radii $R$ is an especially simple example of the
compact internal space of extra dimensions $K_d$. The eigenfunctions
and eigenvalues in this special case look like 
\be
Y_n(y) = \frac{1}{\sqrt{V_d}} \exp \left(i \sum_{m=1}^{d}
n_{m}y^{m}/R
\right), \label{T}
\ee
\[
\lambda_n = |n|^2,\quad |n|^2= n_1^2 + n_2^2 + \ldots n_d^2, \quad
n=(n_1,n_2, \ldots, n_d),\quad -\infty \leq n_m \leq \infty,
\]
where $n_m$ are integer numbers, $V_d = (2\pi R)^d$ is the
volume of the torus.

To reduce the multidimensional theory to the effective
four-dimensional one we wright a harmonic expansion for
the multidimensional field $\Phi(z)$ 
\be
\Phi(z) = \Phi(x,y) = \sum_{n} \phi^{(n)}(x) Y_{n}(y). 
\label{H}
\ee
The coefficients $\phi^{(n)}(x)$ of the harmonic expansion
(\ref{H}) are called Kaluza-Klein (KK) excitations or KK modes, and
they usually include the zero-mode $\phi^{(0)}(x)$, corresponding to
$n=0$ and the eigenvalue $\lambda_{0} = 0$. Substitution of the KK
mode expansion into action (\ref{S}) and integration over the
internal space $K_{d}$ gives
\be
S = \int d^{4}x \sqrt{-g} \left\{  
\frac{1}{2} \left( \partial_{\mu} \phi^{(0)} \right)^{2} -
\frac{m^{2}}{2}
(\phi^{(0)})^{2} \right. + \frac{g}{4!} (\phi^{(0)})^{4} +
\ee
\[
+\left. \sum_{n \neq 0} \left[\frac{1}{2}
\left(\partial_{\mu} \phi^{(n)}
\right) 
\left(\partial^{\mu} \phi^{(n)} \right)^{*} -\frac 
{m_n^2}{2} \phi^{(n)}\phi^{(n)*} \right] 
+ \frac{g}{4!} (\phi^{(0)})^{2} \sum_{n\neq 0} \phi^{(n)}
\phi^{(n)*}\right\} + \ldots.  
\]
For the masses of the KK modes one obtains 
\be
m_{n}^{2} = m^{2} + \frac{\lambda_{n}}{R^{2}}, \label{m}
\ee
and the coupling constant $g$ of the four-dimensional theory is
related  to the coupling constant $G_{(4+d)}$ of the initial
multidimensional theory by the equation 
\be
  g = \frac{G_{(4+d)}}{V_d},  \label{g}
\ee
where $V_d$ is the volume of the compact internal space of extra
dimensions $K_d$. The fundamental coupling constant $G_{(4+d)}$ has
dimension $[\mbox{mass}]^{-d}$. So, the four-dimensional coupling
constant $g$ is dimensionless one as it should be.
Eqs.~(\ref{m},\ref{g}) represent the basic
relations of Kaluza-Klein scenario.  Similar relations take place for
other types of multidimensional quantum field theoretical models.
From four-dimensional point of view we can interpret each KK mode as
a particle with the mass $m_n$ given by Eq.~(\ref{m}). We see that in
according with Kaluza-Klein scenario any multidimensional
field contains an infinite set of KK modes, i.e. an infinite set of
four-dimensional particles with increasing masses, which is called
the Kaluza-Klein tower. Therefore, an experimental observation of
series KK excitations with a characteristic spectrum of the form
(\ref{m}) would be an evidence of the existence of extra dimensions.
So far the KK partners of the particles of the Standard Model have
not been observed. In the Kaluza-Klein scenario this fact can be
explained by a microscopic small size $R$ of extra dimensions
($R<10^{-17}\,cm$); in that case the KK excitations may be produced
only at super-high energies of the scale $E\sim 1/R > 1\,TeV$. Below
this scale only homogeneous zero modes with $n=0$ are accessible ones
for an observation in recent high energy experiments. That is why,
there is a hope to search the KK excitations at the future LHC and
other colliders. 

However, as we have mentioned above, the solution of hierarchy
problem motivated vice versa to introduce the internal extra space
with a large size ($R\sim 1\,mm$). To resolve this obvious
contradiction recently it has been proposed a remarkable idea of
``brane world picture" according to which all matter fields (except
gravity) are localized on a three-dimensional submanifold -- brane --
embedded in fundamental multidimensional space. In the brane world
scenario extra dimensions may have large and even very large size;
they may also have experimentally observable effects. Even though the
models with the brane world scenario may rather seem as exotic ones,
nevertheless, they provide a base for a nontrivial phenomenological
issues related to the fundamental problems in particle physics and
cosmology. We refer with a pleasure the interested reader to the
excellent review articles \cite{1,2} and many references therein.

In this note we would like to argue in favour of that the extra
dimensions have been observed for a long time in the experiments at
very low energies where the nucleon-nucleon dynamics has been
studied. In this respect, we would like to show that the structure of
proton-proton total cross section at very low energies has a clear
signature of the existence of the extra dimensions.  

\section{Global structure of proton-proton total cross
section}

Recently a simple theoretical formula describing the global structure
of proton-proton total cross-section in the whole range of
energies available up today has been derived. The fit to the
experimental data with the formula was made, and it was shown that
there is a very good correspondence of the theoretical formula to the
existing experimental data obtained at the accelerators \cite{3,4}. 

Moreover it turned out there is a very good correspondence of the
theory to all existing cosmic ray experimental data as well. The
predicted values for  $\sigma_{tot}^{pp}$ obtained
from theoretical description of all existing accelerators data are
completely compatible with the values obtained from cosmic ray
experiments \cite{5}. The global structure of proton-proton total
cross section is shown in Fig. 1 extracted from paper \cite{5}.

\begin{figure}[t]
\begin{center}
\begin{picture}(288,200)
\put(15,10){\includegraphics{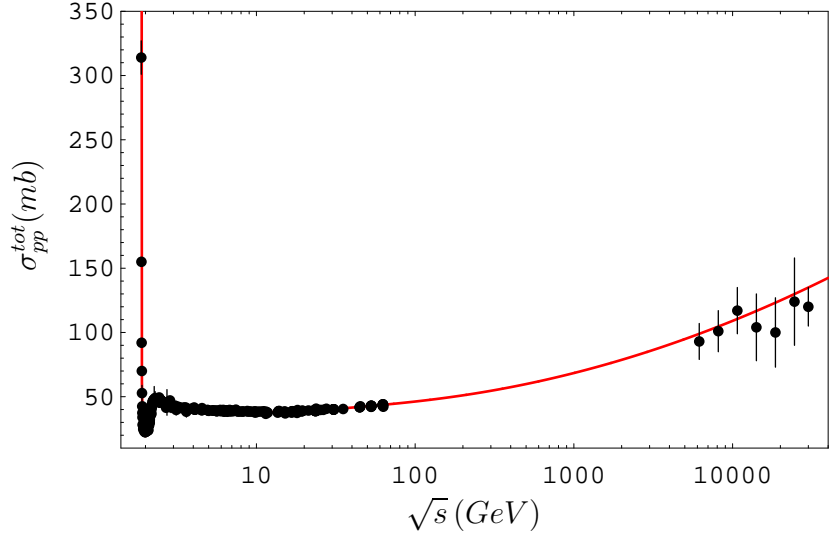}}
\put(144,0){$\sqrt{s}\, (GeV)$}
\put(-5,95){\rotatebox{90}{$\sigma^{tot}_{pp} (mb)$}}
\end{picture}
\caption{\protect{The proton-proton total cross-section versus
$\sqrt{s}$ with the cosmic rays data points from Akeno Observatory
and Fly's Eye Collaboration. Solid line corresponds to our theory
predictions.}}
\label{fig:Fig.1}
\end{center}
\end{figure}

The theoretical formula describing the global structure of
proton-proton total cross section has the following form

\[
\sigma_{pp}^{tot}(s) = \sigma^{tot}_{as}(s) 
\left[1 + \left(\frac{c_1}{\sqrt{s-4m^2_N}R^3_0(s)} -
\frac{c_2}{\sqrt{s-s_{thr}}R^3_0(s)}\right)\left(1 +
d(s)\right)\vert_{s>s_{thr}} +
Res(s)\right],
\]
\[
R^2_0(s) = \left[0.40874044 \sigma^{tot}_{as}(s)(mb) -
B(s)\right](GeV^{-2}),
\]
\[
\sigma^{tot}_{as}(s) = 42.0479 + 1.7548 \ln^2(\sqrt{s}/20.74),
\]
\[
B(s) = 11.92 + 0.3036 \ln^2(\sqrt{s}/20.74),
\]
\[
c_1 = (192.85\pm 1.68)GeV^{-2},\quad c_2 = (186.02\pm 1.67)GeV^{-2},
\]
\[
s_{thr} = (3.5283\pm 0.0052)GeV^2,
\]
\[
d(s) = \sum_{k=1}^{8}\frac{d_k}{s^{k/2}},\qquad Res(s) =
\sum_{i=1}^{N}\frac{C_R^i s_R^i
{\Gamma_R^i}^2}{\sqrt{s(s-4m_N^2)}[(s-s_R^i)^2+s_R^i{\Gamma_R^i}^2]}.
\]
For the numerical values of the parameters $d_i (i=1,...8)$ see
original paper \cite{4}. The mathematical structure of the formula is
very simple and physically transparent: the total cross section is
represented in a factorized form; one factor describes high energy
asymptotics of total cross section and it has the universal energy
dependence predicted by the general theorems in local Quantum Field
Theory (Froissart theorem); the other factor is responsible for the
behaviour of total cross section at low energies and it has a
complicated resonance structure. The nontrivial feature of the
formula is the presence of the new ``threshold" $s_{thr}=3.5283\,
GeV^2$ which is near the elastic one. 

Some information concerning the diproton resonances is collected in
Table 1. The positions of resonances and their widths, listed in
Table 1, were fixed in our fit, and only relative contributions of
the resonances $C_R^i$ have been considered as free fit parameters.
Fitted parameters $C_R^i$ obtained by the fit are listed  in Table 1
too. 
\begin{center}
Table 1. Diproton resonances.

\vspace{5mm}
\begin{tabular}{|l|c|c|r|}\hline   
$ m_R(MeV) $ & $\Gamma_R(MeV) $ & \mbox{Refs.} & $C_R(GeV^2)$  \\
\hline     
$ 1937\pm 2 $ & $ 7\pm 2 $ & $\cite{6}$ & $ 0.058\pm 0.018 $ \\ 
$ 1947(5)\pm 2.5 $ & $ 8\pm 3.9 $ & $\cite{7}$ & $ 0.093\pm 0.028 $\\ 
$ 1955\pm 2 $ & $ 9\pm 4 $ & $\cite{6}$ & $ 0.158 \pm 0.024 $ \\
$ 1965\pm 2 $ & $ 6\pm 2 $ & $\cite{6}$ & $ 0.138 \pm 0.009 $ \\ 
$ 1980\pm 2 $ & $ 9\pm 2 $ & $\cite{6}$ & $ 0.310 \pm 0.051 $ \\
$ 1999\pm 2 $ & $ 9\pm 4 $ & $\cite{6}$ & $ 0.188\pm 0.070 $ \\ 
$ 2008\pm 3 $ & $ 4\pm 2 $ & $\cite{6}$ & $ 0.176 \pm 0.050 $ \\ 
$ 2027\pm ? $ & $ 10 - 12 $ &$\cite{8}$ & $ 0.121\pm 0.018 $ \\ 
$ 2087\pm 3 $ & $ 12\pm 7 $ & $\cite{6}$ & $ -0.069\pm 0.010 $ \\ 
$ 2106\pm 2 $ & $11\pm 5 $ & $\cite{6}$ & $-0.232 \pm 0.025 $ \\ 
$ 2127(9)\pm 5 $ & $ 4\pm 2 $ & $\cite{6}$ & $ -0.222\pm 0.056 $ \\ 
$ 2180(72)\pm 5 $ & $ 7\pm 3 $ & $\cite{6}$ & $ 0.131\pm 0.015 $ \\ 
$ 2217\pm ? $ & $ 8 - 10 $ &$\cite{8}$ & $ 0.112\pm 0.031 $ \\ 
$ 2238\pm 3 $ & $22\pm 8 $ & $\cite{6}$ & $ 0.221 \pm 0.078 $ \\ 
$ 2282\pm 4 $ & $24\pm 9 $ & $\cite{6}$ & $ 0.098 \pm 0.024 $ \\
\hline
\end{tabular}
\end{center}
\vspace{5mm}
Our fitting curve concerning low-energy region is shown in Fig.~2. We
also plotted in Fig.~3 the
resonance structure of proton-proton total cross section at low
energies without the experimental points but with dashed line
corresponding the ``background" where all resonances are switched
off. As it is seen from this Figure there is a clear signature for
the diproton resonances. We may conclude that the diproton resonances
are confirmed by the data set for proton-proton total cross section
at low energies from statistical point of view by the good fit
\cite{8}.

From the global structure of proton-proton total cross-section it
follows that the new ``threshold", which is near the elastic one,
looks like a manifestation of a new unknown particle:
\be
\sqrt{s_{thr}} = 2 m_p + m_{\cal L},\qquad m_{\cal L} =
1.833\,MeV.\label{L}
\ee
This particle was called as $\cal L$-particle from the word {\it
lightest}. It should be emphasized that we predicted the position of
the new ``threshold" with a high accuracy. Of course, the natural
questions have been arisen.  What is the physical nature and
dynamical origin of $\cal L$-particle? Could $\cal L$-particle be
related to the experimentally observed diproton resonances spectrum?
In the next section we'll give the answers to these questions.

\section{$\cal L$-particle in Kaluza-Klein world}

Here we would like to apply the main issues of Kaluza-Klein approach
to our concrete case. Let us assume that $\cal L$-particle is related
to the first KK excitation in the diproton system. Using formula
(\ref{m}) for the masses of KK modes, we can calculate the scale
(size) $R$ of the compact internal extra space. So, starting from the
formula
\be
\sqrt{s_{thr}}=2m_p+m_{\cal L}=2\sqrt{m_p^2+\frac{1}{R^2}},\label{RL}
\ee

\newpage

\begin{figure}[htb]
\begin{center}
\begin{picture}(288,188)
\put(15,10){\includegraphics{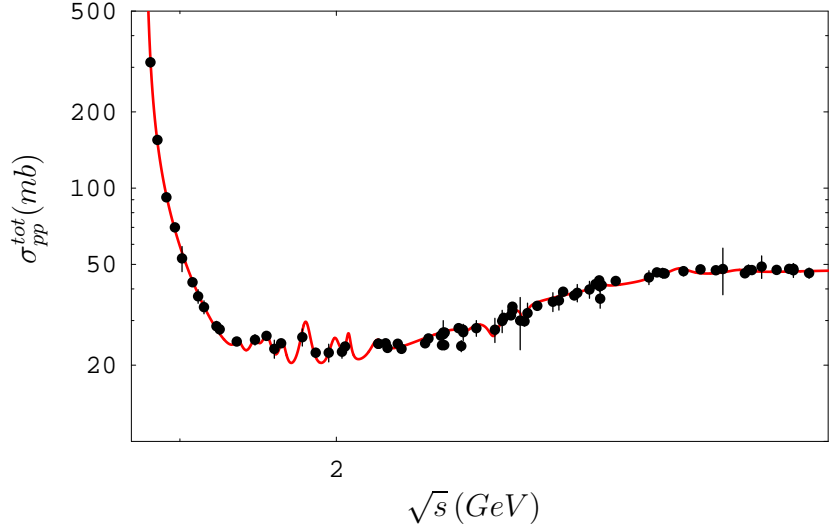}}
\put(144,0){$\sqrt{s}\, (GeV)$}
\put(-5,95){\rotatebox{90}{$\sigma^{tot}_{pp} (mb)$}}
\end{picture}
\caption{\protect{The proton-proton total cross-section versus
$\sqrt{s}$ at low energies. Solid line corresponds to our theory
predictions.}}
\label{fig:Fig.2}
\end{center}
\end{figure}

\vspace{1cm}

\begin{figure}[htb]
\begin{center}
\begin{picture}(288,188)
\put(15,10){\includegraphics{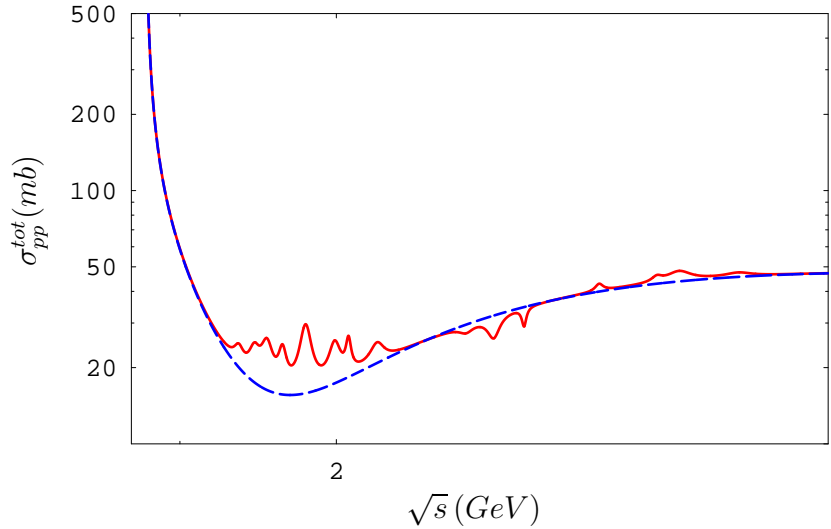}}
\put(144,0){$\sqrt{s}\, (GeV)$}
\put(-5,95){\rotatebox{90}{$\sigma^{tot}_{pp} (mb)$}}
\end{picture}
\caption{\protect{The resonance structure for the proton-proton
total cross-section versus $\sqrt{s}$ at low energies. Solid line is
our theory predictions. Dashed line corresponds to the ``background"
where all resonances are switched off.}}
\label{fig:Fig.3}
\end{center}
\end{figure}

\noindent one obtains
\be
\frac{1}{R}=\sqrt{m_{\cal L}(m_p+\frac{1}{4}m_{\cal L})}=41.481\,
MeV,\label{scale}
\ee
where $m_p=938.272\,MeV$ for the proton mass and Eq.~(\ref{L}) for
the
mass of $\cal L$-particle have been used. From Eq.~(\ref{scale}) it
follows
\be
R=24.1\,GeV^{-1}=4.75\,10^{-13}\mbox{cm}.\label{size}
\ee
It should be emphasized a remarkable fact: the size (\ref{size}) just
corresponds to the scale of distances where the strong Yukawa forces
in strength come down to the electromagnetic forces
$$g_{eff}=g_{\pi NN}\exp(-m_{\pi}R)\sim 0.5,\ \ \  (g^2_{\pi
NN}/4\pi=14.6).$$
On the other hand, for the fundamental mass scale calculated by
formulae (\ref{M}) or (\ref{M2}) with account of size (\ref{size}) in
the case $d=6$  we find
\be
M\sim R^{-1}\left(\frac{M_{Pl}}{R^{-1}}\right)^{2/(d+2)}\mid_{d=6}\,
\sim 5\,\mbox{TeV}.\label{SM}
\ee
Mass scale (\ref{SM}) is just the scale accepted in the Standard
Model, and this is an interesting observation as well.

Going further on, let us build the Kaluza-Klein tower of KK
excitations by the formula 
\be
M_n=2\sqrt{m_p^2+\frac{n^2}{R^2}},\quad (n=1,2,3,\ldots)\label{KK}
\ee
and compare it with the observed irregularities in the spectrum of
mass of the diproton system.\footnote{Similar formula has been
discussed in the literature \cite{17} but with a different physical
interpretation.} The result of the comparison is shown in
Table 2. As it is seen from the Table 2, there is a quite remarkable
correspondence of the Kaluza-Klein picture with the
experiment.

Now, let us suppose that effective bosons $B_n$ with the masses
$m_n=n/R$ related to KK-excitations of a proton may have an effective
Yukawa-type interaction with the fermions
\be
L_{eff} = g_{eff}\bar\psi_f O\psi_f B_n,\label{Leff}
\ee
where $f$ denotes some fermion, for example lepton or quark. If
$m_n>2m_f$ then effective bosons may decay into fermion-antifrermion
pair. For the partial width of such decay in the lowest order over
coupling constant we have
\be
\Gamma_n = \frac{\alpha_{eff}m_n}{2}F_O(x^2_n),\label{Gamma}
\ee
where $\alpha_{eff}=g^2_{eff}/4\pi$, $x^2_n=m^2_f/m^2_n$,
$F_O(x^2)=(1-4x^2)^{3/2}$ for $O=1$ and $F_O(x^2)=(1-4x^2)^{1/2}$ for
$O=\gamma_5$. In that case one obtains an estimation
\be
\Gamma_n \sim n\cdot 0.4\,\mbox{MeV}.\label{Gamma2}
\ee

It's clear from the physics under consideration that a life time of
the diproton resonances will be defined by the decays of effective
bosons $B_n$. This is a remarkable fact that crude estimation
(\ref{Gamma2}) is in a good agreement with an experiment and gives an
explanation of (super)narrowness of dibaryons peaks. Moreover,
estimation (\ref{Gamma2}) shows that the larger the dibaryon mass is,
the larger is the width of the dibaryon.  

Certainly, we have considered here the simplest case of
Kaluza-Klein picture: The built KK-tower corresponds to either
one-dimensional compact extra space or d-dimensional equal radii
torus with the constraint
\be
n = \sqrt{n_1^2 + n_2^2 + \ldots n_d^2}=1,2,3,\ldots,\label{Dio}
\ee
where $n_i(i=1,\ldots,d)$ are integer numbers.

\newpage
\begin{center}
Table 2. Kaluza-Klein tower of KK excitations of diproton system.

\vspace{5mm}
\begin{tabular}{|c|c|lc|}   \hline
n & $M_n(MeV)$ & $M_{exp}^{pp}(MeV)$ & Refs.\\ \hline
1 & 1878.38 & 1877.5 $\pm$ 0.5 & [9] \\ \hline
2 & 1883.87 & 1886 $\pm$ 1 & [6] \\ \hline
3 & 1892.98 & 1898 $\pm$ 1 & [6] \\ \hline
4 & 1905.66 & 1904 $\pm$ 2 & [10] \\ \hline 
5 & 1921.84 & 1916 $\pm$ 2 & [6] \\
  &         & 1926 $\pm$ 2 & [10] \\ \hline
  &         & 1937 $\pm$ 2 & [6] \\
6 & 1941.44 & 1942 $\pm$ 2 & [10] \\
  &  & $\sim$1945 & [7] \\ \hline
7 & 1964.35 & 1965 $\pm$ 2 & [6] \\
  &         & 1969 $\pm$ 2 & [11] \\ \hline 
8 & 1990.46 & 1980 $\pm$ 2 & [6] \\
  &         & 1999 $\pm$ 2 & [6] \\ \hline
9 & 2019.63 & 2017 $\pm$ 3 & [6] \\ \hline
  &         & 2035 $\pm$ 8 & [16] \\
10 & 2051.75 & 2046 $\pm$ 3 & [6] \\
   &         & $\sim$2050 & [12] \\ \hline
11 & 2086.68 & 2087 $\pm$ 3 & [6] \\ \hline
   &         & $\sim$2122 & [12] \\
12 & 2124.27 & 2121 $\pm$ 3 & [13] \\
   &         & 2129 $\pm$ 5 & [6] \\ \hline
   &         & 2140 $\pm$ 9 & [16] \\
13 & 2164.39 & $\sim$2150 & [12] \\
   &         & 2172 $\pm$ 5 & [6] \\ \hline
   &         & 2192 $\pm$ 3 & [13] \\
14 & 2206.91 & 2217 & [8] \\
   &         & 2220 & [15] \\ \hline
15 & 2251.67 & 2238 $\pm$ 3 & [6] \\   
   &         & 2240 $\pm$ 5 & [13] \\ \hline
16 & 2298.57 & 2282 $\pm$ 4 & [6,14] \\ \hline
17 & 2347.45 & 2350 & [15] \\ \hline
\end{tabular}
\end{center}
The constraint
(\ref{Dio}) corresponds to the special (Diophantus!) selection of the
states. It's clear that in
general case of generic extra compact manifold we would have a
significantly more wealthy spectrum of KK-excitations. We could
imagine that there exist such extra compact manifold with a suitable
geometry where KK-excitations of a few input fundamental entities
(proton, electron, photon, etc.) would provide the experimentally
observed spectrum of all particles, their resonances and nuclei
states. As we hope, it would be possible to find in this way
the {\it global} solution of the Spectral Problem. Anyhow, we believe
that such perfect extra compact manifold with a beautiful geometry
and its good-looking shapes exist.

\section{Conclusion}

In this short article we have presented some arguments in favour of
Kaluza and Klein ideas genius which are waiting their time in
high-energy experiments at future colliders. In fact, we have shown
that Kaluza-Klein picture of the world has been confirmed in the
experiments  at very low energies where the nucleon-nucleon dynamics
has been studied. Geniusly simple formula (\ref{KK}) provided by
Kaluza-Klein approach so accurately describes the mass spectrum of
diproton system, it's very nice; certainly, it is not an accidental
coincidence. 

Here we have concerned the simplest model where
the protons were considered as a scalar particles. It is well known
that account of fermionic degrees of freedom may result the
nontrivial problems related to both the index and the kernel of Dirac
operator on a generic compact manifold. However, since the kernel of
Dirac operator is equal to the kernel of its square, we can say with
confidence that account of fermionic degrees of freedom for a proton
will not change our main conclusion. This conclusion is that the
existence of the extra dimensions was experimentally proved for a
long time, but we did not understand it. Now, seems we understand it.


\begin{thebibliography}{**}
\bibitem{1}
V.A.~Rubakov, {\it Large and infinite extra dimensions}, Sov. J.
Uspekhi {\bf 171}, 913 (2001); e-print hep-ph/0104152.
\bibitem{2}
Yu.A.~Kubyshin, e-print hep-ph/0111027.
\bibitem{3}
A.A.~Arkhipov, \emph{What Can we Learn from the Study of Single
Diffractive
Dissociation at High Energies?} -- in Proceedings of VIIIth Blois
Workshop on Elastic and Diffractive Scattering, Protvino, Russia,
June 28--July 2, 1999, World Scientific, Singapore, 2000,
pp.~109-118; preprint IHEP 99-43, Protvino, 1999; e-print
hep-ph/9909531. 
\bibitem{4}
A.A.~Arkhipov, \emph{On Global Structure of Hadronic Total Cross
Sections},
preprint IHEP 99-45, Protvino, 1999; e-print hep-ph/9911533.
\bibitem{5}
A.A.~Arkhipov, \emph{Proton-Proton Total Cross Sections from the
Window of
Cosmic Ray Experiments}, preprint IHEP 2001-23, Protvino, 2001;
e-print hep-ph/0108118; in Proceedings of IXth
Blois Workshop on Elastic and Diffractive Scattering, Pruhonice near
Prague, Czech Republic, June 9-15, 2001, eds. V.~Kundrat, P.~Zavada,
Institute of Physics, Prague, Czech Republic, 2002, pp.~293-304.
\bibitem{6}
Yu.A.~Troyan, V.N.~Pechenov, Sov. J. Yad. Phys. {\bf 56}, 191 (1993); 
Yu.A.~Troyan, Sov. J. Physics of Element. Part. and Atomic Nuclei
{\bf 24}, 683 (1993).
\bibitem{7}
B.~Tatischeff et al., Phys. Rev. C{\bf 45} 2005 (1992).
\bibitem{8}
A.A.~Arkhipov, \emph{On a Manifestation of Dibaryon Resonances in the
Structure of Proton-Proton Total Cross Section at Low Energies},
preprint IHEP 2001-44, Protvino, 2001;
e-print hep-ph/0110399; in Proceedings of the Ninth International
Conference on Hadron Spectroscopy, Protvino, Russia, 25 August-1
September, 2001, eds. D.~Amelin, A.M.~Zaitsev, Melville, New York,
2002, AIP Conference Proceedings, vol. 619, pp. 771-776. 
\bibitem{9}
B.M.~Abramov et al., Sov. J. Yad. Phys. {\bf 57}, 850 (1994).
\bibitem{10}
L.V.~Filkov et al., e-prints nucl-th/0009044; hep-ex/0006029.
\bibitem{11}
B.~Tatischeff et al., Z. Phys. A{\bf 328} 147 (1987).
\bibitem{12}
B.~Tatischeff et al., Phys. Rev. C{\bf 59} 1878 (1999).
\bibitem{13}
B.~Tatischeff et al., Phys. Rev. C{\bf 36} 1995 (1987).
\bibitem{14}
B.~Tatischeff et al., in Proceedings of the Xth International Seminar
on High Energy Physics Problems, Dubna, 24-29 September, 1990, USSR,
eds. A.M.~Baldin, V.V.~Burov, L.P.~Kaptari, World Scientific, 1990,
p. 177; in Proceedings of the XIIth International Seminar
on High Energy Physics Problems, Vol. II, Dubna, 1997, p.~62. 
\bibitem{15}
Y.~Ohashi et al., Phys. Rev. C{\bf 36} 2432 (1987).
\bibitem{16}
V.P.~Andreev et al., Z. Phys. A{\bf 327} 363 (1987).
\bibitem{17}
F.A.~Gareev, G.S.~Kazacha, Yu.L.~Ratis, Sov. J. Physics of Element.
Part. and Atomic Nuclei
{\bf 27}, 97 (1996).

\end{thebibliography}
\end{document}